\documentclass[a4paper,iop,twocolumn,twocolappendix,numberedappendix,appendixfloats]{emulateapj}
\pdfoutput=1
\usepackage[breaklinks,colorlinks=true,linkcolor=blue,anchorcolor=blue,citecolor=blue,urlcolor=blue,draft=false]{hyperref}
\usepackage{color}
\usepackage{graphicx}
\usepackage{amsmath} 
\usepackage{amssymb}
\usepackage{mathrsfs}
\usepackage{placeins}
\usepackage{times}
\usepackage{xspace}

\usepackage{natbib}

\newcommand{\percent}{\ensuremath{\%}}
\newcommand{\simgt}{\lower.5ex\hbox{$\; \buildrel > \over \sim \;$}}
\newcommand{\simlt}{\lower.5ex\hbox{$\; \buildrel < \over \sim \;$}}

\newcommand{\Mtot}{M_\mathrm{tot}}
\newcommand{\Mgas}{M_\mathrm{gas}}
\newcommand{\Mbar}{M_\mathrm{bar}}
\newcommand{\Mstar}{M_\mathrm{star}}
\newcommand{\gbar}{g_\mathrm{bar}}
\newcommand{\gtot}{g_\mathrm{tot}}
\newcommand{\fgas}{f_\mathrm{gas}}
\newcommand{\fbar}{f_\mathrm{bar}}
\newcommand{\fcold}{f_\mathrm{c}}

\lefthead{Tian et al.}
\righthead{RAR in CLASH Clusters}

\begin{document}
\nocite{*}

\title{The Radial Acceleration Relation in CLASH Galaxy Clusters}
\author{Yong Tian\altaffilmark{1}}
\author{Keiichi Umetsu\altaffilmark{2}}
\author{Chung-Ming Ko\altaffilmark{1,3}}
\author{Megan Donahue\altaffilmark{4}}
\author{I-Non Chiu\altaffilmark{2}}

\email{yongtian@gm.astro.ncu.edu.tw}
\email{keiichi@asiaa.sinica.edu.tw}
\email{cmko@gm.astro.ncu.edu.tw}
\altaffiltext{1}
 {Institute of Astronomy, National Central University, Taoyuan 32001,
 Taiwan}
\altaffiltext{2}
 {Academia Sinica Institute of Astronomy and Astrophysics (ASIAA),
 No. 1, Section 4, Roosevelt Road, Taipei 10617, Taiwan}
\altaffiltext{3}
 {Department of Physics and Center for Complex Systems, National Central University, Taoyuan 32001, Taiwan}
\altaffiltext{4}
 {Physics and Astronomy Department, Michigan State University, East Lansing, MI 48824, USA}

\begin{abstract}
The radial acceleration relation (RAR) in galaxies describes a tight
 empirical scaling law between the total acceleration
 $\gtot(r)=G\Mtot(<r)/r^2$
 observed in galaxies and that expected from their baryonic mass
 $\gbar(r)=G\Mbar(<r)/r^2$,
 with a characteristic acceleration scale of
 $g_\dag\simeq 1.2\times 10^{-10}$\,m\,s$^{-2}$.
Here, we examine if such a correlation exists in galaxy clusters using
 weak-lensing, strong-lensing, and X-ray data sets available for 20
 high-mass clusters targeted by the CLASH survey.
By combining our CLASH data with stellar mass estimates for the
 brightest cluster galaxies (BCGs) and accounting for the stellar
 baryonic component in clusters, we determine, for the first time,
 an RAR on BCG--cluster scales.
The resulting RAR is well described by a tight power-law relation,
 $\gtot\propto \gbar^{0.51^{+0.04}_{-0.05}}$, with lognormal intrinsic
 scatter of $14.7^{+2.9}_{-2.8}\percent$.
The slope is consistent with the low acceleration limit of the
 RAR in galaxies, $\gtot=\sqrt{g_\dag\,\gbar}$,
 whereas the intercept implies a much higher acceleration scale
 of $g_\ddag = (2.02\pm0.11)\times 10^{-9}$\,m\,s$^{-2}$,
 indicating that there is no universal RAR that holds on all scales from
 galaxies to clusters.
We find that the observed RAR in CLASH clusters is consistent with
 predictions from a semi-analytical model developed in the
 standard $\Lambda$CDM framework.
Our results also predict the presence of a baryonic Faber--Jackson
 relation ($\sigma_v^4\propto M_\mathrm{bar}$) on cluster
 scales.
\end{abstract}

\section{Introduction}\label{sec:intro}

Clusters of galaxies exhibit a large mass discrepancy between baryonic and
gravitational mass. Understanding the nature and amount of unseen mass
in galaxy clusters is a long-standing issue in astrophysics.
\citet{Zwicky33} was the first to analyze dynamics of the Coma cluster and
to infer the existence of ``dark matter'' (DM).
The vast majority of baryons (80--90\%) in galaxy clusters are in the form of X-ray
emitting diffuse hot gas. The gas mass fraction in high-mass galaxy clusters
is observed to reach $\simeq 13\percent$ in the intracluster region
\citep[e.g.,][]{Vikhlinin2006_fgas,Umetsu09,planck2013fgas,Donahue14}, and
their global mass content is dominated by DM ($\sim 85\percent$).
Moreover, cold dark matter (CDM) that dominates the matter budget of the
universe is essential to explain a range of cosmological probes
on larger scales, such as cosmic microwave background
anisotropy, large-scale galaxy clustering, and weak-lensing
cosmic-shear observations.
The current concordance cosmological paradigm, $\Lambda$CDM,
also assumes a cosmological constant ($\Lambda$) to account for the late-time
accelerated expansion of the universe in the framework of general relativity.

At the scales of spiral galaxies, the discrepancy between baryonic and
    dynamical mass, $\Mtot(<r)/\Mbar(<r)$, is found to tightly couple
    with gravitational acceleration, but no obvious correlation with
    other physical quantities, such as the size and orbital frequency,
    was found so far    \citep[e.g.,][]{McGaugh04}.
    The mass ratio $\Mtot(<r)/\Mbar(<r)$ increases systematically with decreasing
    acceleration below a characteristic sale of
    $\simeq 10^{-10}$\,m\,s$^{-2}$.
    This is referred to as the mass discrepancy--acceleration relation
    \citep[MDAR; for a review, see, e.g.,][]{FM12}.

In the case of spiral galaxies, the low acceleration limit gives a
    baryonic Tully--Fisher relation and an acceleration scale that is
    consistent with the MDAR \citep{McGaugh11}.
    Using a sample of 153 disk galaxies from the SPARC database \citep{Lelli16},
    \citet{McGaugh16} found a tight radial acceleration relation (RAR)
    between the observed total acceleration $\gtot=G\Mtot(<r)/r^2$
    and the baryonic acceleration
    $\gbar=G\Mbar(<r)/r^2$
    defined at the same galacto-centric radius $r$ as
    \begin{equation}
     \label{eq:MDAR}
      \frac{\gtot}{\gbar}=\frac{\Mtot}{\Mbar}=\frac{1}{1-e^{-\sqrt{\gbar/g_{\dag}}}}\,,
    \end{equation}
where
$g_{\dag}=1.20\pm0.02\,(\mathrm{stat.})\pm0.24\,(\mathrm{syst.})\times10^{-10}$\,m\,s$^{-2}$.
The low acceleration limit ($\gbar\ll g_\dag$) of Equation
(\ref{eq:MDAR}) gives the following power-law relation:
\begin{equation}
 \label{eq:MDARlow}
 \gtot = \sqrt{g_\dag\,\gbar}.
\end{equation}
Similarly, the MDAR \citep{Scarpa06, Janz16, TK16} and
the RAR observed in elliptical galaxies are consistent with those
of spiral galaxies \citep{Lelli17, Rong18, Chae19, Milgrom19, TK19}.
Moreover, the MDAR of 53 elliptical galaxies obtained from strong-lens
    modeling of Einstein rings is consistent with the dynamical results
    \citep{TK17}.

The RAR observed on galaxy scales raises four issues to be addressed
    \citep{Desmond17, Lelli17}:
    (1) the characteristic acceleration scale $g_{\dag}$;
    (2) the slope in the small acceleration limit ($\simeq 0.5$);
    (3) the tightness of intrinsic scatter of the relation (0.11 dex);
    (4) no correlation with other galactic properties.
Possible explanations of the above issues can be classified into the
    following three categories \citep{McGaugh16, Lelli17}:
    (I) galaxy formation processes in the ${\Lambda}$CDM model;
    (II) new ``dark sector'' physics; and
    (III) new dynamical laws.

    In the ${\Lambda}$CDM framework, baryonic mass dominated by stars is
    bound to the gravitational potential well of the DM halo
    hosting a galaxy.
    However, the baryons are re-distributed through complex galaxy formation
    processes, such as active galactic nuclei (AGN) feedback,
    stellar winds, and supernova explosions.
    Several attempts have been made to explain the observed RAR in the
    context of $\Lambda$CDM through hydrodynamical simulations
    \citep{WK15, Ludlow17}
    and

    semi-empirical models of galaxy formation \citep{DL16, Navarro17, Desmond17}
    by adopting an abundance matching relation \citep{Behroozi13}.
    Besides the standard CDM paradigm,
    some considered baryon-DM coupling in a dark-fluid framework \citep{ZL10,
    Khoury15} or dipolar DM particles \citep{BL08, BL09}.

    Alternatively, the RAR has been interpreted as a consequence of a
    new dynamical law without the need of DM.
    \cite{Milgrom83} introduced modified Newtonian dynamics (MOND),
    in which the dynamical law changes with an acceleration scale $g_{\dag}$.
    However, MOND cannot explain the dynamics of galaxy clusters, and it
    needs to account for a large missing mass of about a factor of two \citep{PS05, Sanders99, Sanders03}.

    The four issues posed by the RAR observed in galaxies also
    represent some challenges to the ${\Lambda}$CDM model
    \citep{Desmond17, Lelli17}.
    \citet{WK15} found that simulation results do not match well the
    data presented in  \cite{McGaugh04}.
    \citet{Ludlow17} successfully reproduced (2) and (4) within the
    $\Lambda$CDM framework, but with an acceleration scale $2.2g_{\dag}$,
    which is significantly higher than the observed value \citep{Li18}.
    \citet{DL16} explained (1) and (2) by introducing an abundance matching prescription,
    whereas the level of scatter is significantly larger than the
    observed value, and the residuals are correlated with the
    galacto-centric radius.
    \citet{Desmond17} used more sophisticated model but got still larger
    scatter even using zero scatter in abundance matching.

    How about galaxy clusters? \citet{Navarro17} argued in the
    $\Lambda$CDM framework that
    the RAR in galaxy clusters (if exists) should be deviated
    from that of galaxies because the
    central maximum halo acceleration exceeds $g_{\dag}$
    ($\simeq3\times10^{-10}$\,m\,s$^{-2}$).
    On the other hand, MOND predicts the RAR in galactic systems.
    Its failure in galaxy clusters would pose a significant challenge to
    MOND \citep{FM12}.
    Although the RAR is well studied in galactic systems, it
    has never been explored in galaxy clusters.

In this paper, we present an RAR in 20 high-mass galaxy clusters based on
high-quality multiwavelength data sets available for the CLASH survey.
The paper is organized as follows.
In Section~\ref{sec:data}, we summarize the characteristics of the CLASH
sample and the data products.
In Section~\ref{sec:results}, we present the radial profiles of the
baryon fraction and the RAR for the CLASH sample.
In Section~\ref{sec:discussion}, we discuss the results and implications of our findings.
Finally a summary is given in Section~\ref{sec:summary}.

Throughout this paper, we assume a flat $\Lambda$CDM cosmology
with
$\Omega_\mathrm{m}=0.27$, $\Omega_\Lambda=0.73$, and a Hubble constant of
$H_0 = 100h$\,km\,s$^{-1}$\,Mpc$^{-1}$
with $h=0.7$.
We denote the critical density of the universe at a particular redshift
$z$ as $\rho_\mathrm{c}(z)=3H^2(z)/(8\pi G)$, with $H(z)$ the
redshift-dependent Hubble parameter.
We adopt the standard notation
$M_\Delta$
to denote the mass enclosed within a sphere of radius
$r_\Delta$
within which the mean overdensity equals
$\Delta \times \rho_\mathrm{c}(z)$. That is,
$M_\Delta=(4\pi\Delta/3)\rho_\mathrm{c}(z)r_\Delta^3$.
We define the gravitational acceleration
in the framework of Newtonian dynamics as $g(r)=GM(<r)/r^2$.

\section{Cluster Sample and Data}
\label{sec:data}

We analyze multiwavelength data products from the Cluster Lensing And
Supernova survey with Hubble \citep[CLASH;][]{Postman12}.
The CLASH survey is a 524-orbit {\em Hubble Space Telescope} ({\em HST})
Multi-Cycle Treasury program designed to probe the mass
distribution of 25 high-mass galaxy clusters with
$M_{500}\simgt 4\times 10^{14}M_\odot$ \citep{Umetsu16}.
In this sample, 20 clusters were X-ray selected to be hot ($T_X>5$\,keV)
and to have a regular X-ray morphology.
Numerical simulations
suggest that the X-ray-selected subsample is largely composed of
relaxed clusters ($\sim 70\percent$), but it also contains a nonnegligible
fraction ($\sim 30\percent$) of unrelaxed systems \citep{Meneghetti14}.
Another subset of five clusters were selected by their lensing
properties to produce high-magnification events.
These clusters often turn out to be dynamically disturbed merging
systems.

In this study, we focus on a subset of 20 CLASH clusters taken from
\citet{Umetsu16}, who presented a joint
analysis of strong-lensing, weak-lensing shear and magnification data of
these individual clusters.
Among the 20 clusters, 16 are X-ray selected, and the rest
are high-magnification systems.
The full-lensing analysis of \citet{Umetsu16} combined constraints from
16-band {\em HST}
observations \citep{Zitrin15} and wide-field multicolor imaging taken
primarily with Suprime-Cam on the Subaru telescope \citep{Umetsu14}.
For all clusters in the CLASH sample, \citet{Donahue14} derived binned
radial profiles of temperature, gas mass, and hydrostatic mass using
{\em Chandra} and {\em XMM-Newton} X-ray observations.

Here we combine the total mass measurements $\Mtot$ of \citet{Umetsu16}
based on strong and weak lensing (Section~\ref{subsec:lens})
and the X-ray gas mass measurements $\Mgas$ of \citet{Donahue14}
to study the relationship
between the total and baryonic acceleration profiles.  Moreover, we
statistically account for the stellar contribution to the baryonic
acceleration (Section~\ref{subsec:stars}).  We also include stellar mass
estimates for the central brightest cluster galaxies (BCGs; Section~\ref{subsec:BCG}).

\subsection{Lensing Mass}
\label{subsec:lens}

Combining the strong lensing, weak lensing shear and magnification
effects, \citet{Umetsu16} reconstructed the surface mass density
profile of each individual cluster over a wide range of
the cluster-centric radius.
\citet{Umetsu16} found that
the ensemble-averaged total mass distribution of the CLASH sample is
well described by a family of cuspy, outward-steepening density
profiles, namely,
the Navarro--Frenk--White \citep[NFW, hereafater;][]{Navarro97}, Einasto,
and DARKexp \citep{DARKexp} models.
Of these, the NFW model best describes the CLASH lensing data
\citep[e.g.,][]{Umetsu11,Umetsu17}.
On the other hand, the single power-law, cored isothermal, and Burkert
profiles were statistically disfavored by the observed CLASH lensing
profile having a pronounced radial curvature \citep{Umetsu16}.

Here we use the CLASH lensing
constraints on the total mass profile $\Mtot(<r)$ of each individual
CLASH cluster assuming a spherical NFW profile. The total mass
$\Mtot(<r)$ of an NFW halo as a function of spherical radius $r$ is
written as
\begin{equation}
 \Mtot(<r|M_{200},c_{200}) = 4\pi \rho_\mathrm{s} r_\mathrm{s}^3
  \left[\ln\left(1+\frac{r}{r_\mathrm{s}}\right)-\frac{r}{r+r_\mathrm{s}}\right],
\end{equation}
where $r_\mathrm{s}$ and $\rho_\mathrm{s}$ represent
the characteristic scale radius and density of the NFW profile,
respectively, and $\rho_\mathrm{s}$ is given by
\begin{equation}
 \rho_\mathrm{s} = \frac{200}{3}\frac{c_{200}^3}{\ln(1+c_{200})-c_{200}/(1+c_{200})}\rho_\mathrm{c}(z)
\end{equation}
with $c_{200}\equiv r_{200}/r_\mathrm{s}$ the NFW concentration
parameter.

For each cluster, \citet{Umetsu16} extracted the posterior probability
distributions of $(M_{200},c_{200})$ from the observed surface
mass density profile assuming a spherical NFW halo, by accounting for
all relevant sources of uncertainty
\citep[e.g.,][]{Umetsu16,Umetsu20,Miyatake19}:
(i) measurement errors,
(ii) cosmic noise due to projected large-scale structure uncorrelated
with the cluster,
(iii) statistical fluctuations of the projected cluster lensing signal
due to halo triaxiality and correlated substructures.
According to cosmological hydrodynamical
simulations of \citet{Meneghetti14}, the CLASH sample selection is
expected to be largely free from orientation bias.
In fact, three-dimensional full-triaxial analyses of CLASH lensing
observations found no statistical evidence for orientation bias in the
CLASH sample \citep[see][]{Sereno18,Chiu18}.
Therefore, assuming spherical NFW halos is not expected to cause any
significant bias in lensing mass estimates of the CLASH sample.
The mass and concentration parameters $(M_{200}, c_{200})$ of each
individual CLASH cluster are summarized in Table 2 of
\citet[][]{Umetsu16}.

In our analysis, we use these posterior distributions of the NFW
parameters to obtain well-characterized inference
of $\Mtot(<r|M_{200},c_{200})$ for each individual cluster.
We compute the total mass profile $\Mtot(<r|M_{200},c_{200})$ and its
uncertainty of each cluster in the radial range at $r\ge
r_\mathrm{min}\simeq 14$\,kpc.

\subsection{Baryonic Mass}
\label{subsec:stars}

The X-ray emitting hot gas dominates the baryonic mass in galaxy
clusters. In high-mass clusters, more than $80\percent$ of the
intra-cluster baryons are in the X-ray emitting hot phase
\citep[e.g.,][]{Umetsu09, Donahue14, Okabe14, Chiu18}.
\citet{Donahue14} derived enclosed gas mass profiles $\Mgas(<r)$ for all
CLASH clusters, finding that the $\Mgas$ profiles measured from the {\em
Chandra} and {\em XMM} observations are in excellent agreement
where the data overlap.
Since {\em XMM} data are not available for all the clusters,
we only use {\em Chandra} $\Mgas$ measurements of
\citet{Donahue14} as primary constraints on the baryonic mass content in
the CLASH sample.  For each cluster, we have measured $\Mgas(<r)$ values
in several radial bins, where the radial range is different for each
cluster, depending on the redshift and the data quality (see Figure~\ref{Fig:1}).

We then account for the stellar contribution to the baryonic mass using
the results of \citet{Chiu18}, who established the stellar-to-gas mass
relation $\fcold(r)=\Mstar/(\Mgas+\Mstar)$ (see their Figure~11;
$\fcold$ referred to as the cold collapsed baryonic fraction),
for a sample of 91 Sunyaev--Zel'dovich effect (SZE) clusters with
$M_{500}>2.5\times 10^{14}M_\odot$
selected from
the South Pole Telescope (SPT) survey \citep{carlstrom2011,bleem2015}.
The $\fcold(r)$ relation is insensitive to the cluster redshift over a
broad range out to $z\sim 1.3$ as probed by the SPT sample.
In their study, the total masses were estimated from the SZE
observable, the gas masses $\Mgas$ from {\em Chandra} X-ray data, and
the stellar masses $\Mstar$ from combined optical/near-infrared
multi-band photometry.
The $\fcold(r)$ relation of \citet{Chiu18} includes the stellar mass
contributions from the BCG and cluster member galaxies inside the
$r_{500}$ overdensity radius.

With the mean $\fcold(r)$ relation, we estimate the total baryonic
mass as $\Mbar(<r) = \Mgas(<r)/[1-\fcold(r)]$, ignoring the
cluster-to-cluster scatter.
The level of scatter around the mean
$\fcold(r)$ relation is about $12\percent$.

\subsection{BCG Stellar Mass}
\label{subsec:BCG}

In the central cluster region, the baryonic mass in clusters is
dominated by the stellar mass in the BCG.
In this study, we model the stellar mass distribution of each BCG
with the Hernquist model \citep{Hernquist90},
 which gives an analytical approximation to the deprojected form of de
 Vaucouleurs' profile (or a S\'ersic profile with index $n=4$).
Then, the stellar mass $m_\mathrm{star}(<r)$ inside the spherical radius
$r$ and the stellar gravitational acceleration $g_\mathrm{star}(<r)$ are
expressed as
\begin{equation}
 \label{eq:Hernquist}
  \begin{aligned}
   m_\mathrm{star}(<r) &=\frac{{\cal M}_\mathrm{star}r^2}{(r+r_\mathrm{h})^2},\\
   g_\mathrm{star}(r)&=\frac{G{\cal M}_\mathrm{star}}{(r+r_\mathrm{h})^2},
   \end{aligned}
\end{equation}
where ${\cal M}_\mathrm{star}$ is the total stellar mass of the BCG, and
$r_\mathrm{h}\approx 0.551 R_\mathrm{e}$
is a characteristic scale length of the Hernquist model,
with $R_\mathrm{e}$ the half-light or effective radius of the de Vaucoleurs'
brightness profile.

In this study, we adopt as ${\cal M}_\mathrm{star}$ (see
Table~\ref{tab:BCG}) the stellar mass
estimates of CLASH BCGs from \citet[][their Table 1]{Cooke16}, who performed a
multiwavelength analsyis on a large sample of BCGs by combining UV,
optical, near-infrared, and far-infrared data sets.
Since the measurement errors on BCG stellar masses were not
provided in \citet{Cooke16}, we assume a fractional uncertainty of
$10\percent$ on ${\cal M}_\mathrm{star}$.

We measure the BCG effective radius $R_\mathrm{e}$
from the CLASH {\em HST} imaging using the GALFIT package \citep{Peng10}.
We choose to measure $R_\mathrm{e}$ of each BCG in
the {\em HST} band corresponding to the rest-frame wavelength of $1\mu$m.
The corresponding {\em HST} bands for our sample ($0.187\le z\le 0.686$)
are all in the WFC3/IR coverage (F110W to F160W), as summarized in
Table~\ref{tab:BCG}.
We fit a single S\'ersic
profile to the surface brightness
distribution of the BCG in the
CLASH {\em HST} imaging data.
The initial guess of the BCG position is
based on the rest-frame UV (280\,nm) measurements of \citet{Donahue15}.
In Table~\ref{tab:BCG}, we also list the final source position
($\mathrm{R.A.}, \mathrm{Decl.}$) in J2000 coordinates,
S\'ersic index ($n$), and effective radius ($R_\mathrm{e}$) from our GALFIT modeling.


\begin{deluxetable*}{lcccccccccc}
 \tablecolumns{11}
 \tablewidth{0pt}
 \tabletypesize{\scriptsize}
 \scriptsize
\tablecaption{\label{tab:BCG}
 Properties of BCGs in the CLASH sample}
 \tablehead{
 \multicolumn{1}{c}{Cluster name} &
 \multicolumn{1}{c}{Redshift\tablenotemark{a}} &
 \multicolumn{1}{c}{R.A.\tablenotemark{a}} &
 \multicolumn{1}{c}{Decl.\tablenotemark{a}} &
 \multicolumn{1}{c}{Band\tablenotemark{b}} &
 \multicolumn{1}{c}{$n$\tablenotemark{c}} &
 \multicolumn{1}{c}{$R_\mathrm{e}$\tablenotemark{d}} &
 \multicolumn{1}{c}{$r$\tablenotemark{e}} &
 \multicolumn{1}{c}{$M_\mathrm{star}$\tablenotemark{f}} &
 \multicolumn{1}{c}{$M_\mathrm{gas}$\tablenotemark{g}} &
 \multicolumn{1}{c}{$\Mtot$\tablenotemark{h}}\\
 \colhead{} & &
 \multicolumn{1}{c}{(J2000.0)} &
 \multicolumn{1}{c}{(J2000.0)} & & &
 \multicolumn{1}{c}{(kpc)} &
 \multicolumn{1}{c}{(kpc)} &
 \multicolumn{1}{c}{($10^{11}M_{\odot}$)} &
 \multicolumn{1}{c}{($10^{11}M_{\odot}$)} &
 \multicolumn{1}{c}{($10^{11}M_{\odot}$)}
 }
\startdata
~~Abell 383        & 0.187  &$02:48:03.38$& $-03:31:45.02$& F110W	&	2.34	&	17.0 	$\pm$	0.09 	&	14.3    &	4.45	& 1.78 $\pm$ 0.09 	&	 7.55 $\pm$ 2.23 \\
~~Abell 209        & 0.206  &$01:31:52.55$& $-13:36:40.50$& F125W	&	2.62 	&	22.1 	$\pm$	0.14    &	14.3  	&	4.85	&	-			    &	 3.87 $\pm$ 0.79 \\
~~Abell 2261       & 0.224  &$17:22:27.21$& $+32:07:57.62$& F125W	&	1.74 	&	18.7 	$\pm$	0.08 	&	23.6 	&  12.30	& 0.48 $\pm$ 0.03	&	 6.44 $\pm$ 1.48 \\
~~RX~J2129.7$+$0005 &	0.234 	&$21:29:39.96$& $+00:05:21.17$& F125W	&	2.70 	&	41.4 	$\pm$	0.54 	&	14.3 	&	5.81	& 2.18 $\pm$ 0.07   &	 6.65 $\pm$ 1.91 \\
~~Abell 611         &	0.288 	&$08:00:56.82$& $+36:03:23.63$& F125W	&	2.55 	&	30.4 	$\pm$	0.16 	&	22.2 	&	6.58	& 0.48 $\pm$ 0.03	&	 6.24 $\pm$ 1.81 \\
~~MS2137$-$2353     &	0.313 	&$21:40:15.16$& $-23:39:40.10$& F125W	&	2.35 	&	15.0 	$\pm$	0.04 	&	14.3 	&	3.65	& 2.94 $\pm$ 0.07   &	 3.98 $\pm$ 1.57 \\
~~RX~J2248.7$-$4431 &	0.348 	&$22:48:43.97$& $-44:31:51.14$& F140W	&	2.45 	&	34.5 	$\pm$	0.20 	&	30.3 	&	8.09	& 1.01 $\pm$ 0.03   &	 6.19 $\pm$ 2.16 \\
~~MACS~J1115.9$+$0129 & 0.355 	&$11:15:51.91$& $+01:29:55.00$& F140W	&	3.83 	&	52.9 	$\pm$	0.93 	&	16.2 	&	3.00    & 5.80 $\pm$ 0.19   &	 6.25 $\pm$ 1.53 \\
~~MACS~J1931.8$-$2635 & 0.352 	&$19:31:49.70$& $-26:34:32.22$& F140W	&	3.49 	&	33.2 	$\pm$	0.38 	&	14.3 	&	6.92	& 1.47 $\pm$ 0.02   &	 7.21 $\pm$ 2.90 \\
~~RX~J1532.9$+$3021   & 0.362 	&$15:32:53.78$& $+30:20:59.43$& F140W	&	2.81 	&	21.8 	$\pm$	0.14 	&	14.3 	&	3.34	& 1.13 $\pm$ 0.04   &	 6.80 $\pm$ 4.18 \\
~~MACS~J1720.3$+$3536  & 0.387 	&$17:20:16.75$& $+35:36:26.24$& F140W	&	2.63 	&	17.2 	$\pm$	0.06 	&	23.6 	&	6.59	& 1.15 $\pm$ 0.03   &	 6.83 $\pm$ 2.07 \\
~~MACS~J0416.1$-$2403 & 0.397 	&$04:16:09.15$& $-24:04:02.99$& F140W	&	3.78 	&	56.2 	$\pm$	0.81 	&	14.3 	&	3.14	&	-			    &	 4.22 $\pm$ 0.94 \\
~~MACS~J0429.6$-$0253 & 0.399 	&$04:29:36.00$& $-02:53:06.78$& F140W	&	1.80 	&	29.3 	$\pm$	0.08 	&	17.2 	&  11.90	& 6.71 $\pm$ 0.55 	&	 9.98 $\pm$ 3.40 \\
~~MACS~J1206.2$-$0847 & 0.439 	&$12:06:12.15$& $-08:48:03.32$& F140W	&	3.65 	&	44.8 	$\pm$	0.52 	&	14.3 	&	3.13	&	-			    &	 6.90 $\pm$ 2.07 \\
~~MACS~J0329.7$-$0211 & 0.450 	&$03:29:41.57$& $-02:11:46.33$& F140W	&	2.76 	&	22.9 	$\pm$	0.12 	&	22.2 	&	8.47	& 13.1 $\pm$ 0.41 	&  25.30 $\pm$ 6.50 \\
~~RX~J1347.5$-$1145   & 0.451 	&$13:47:30.61$& $-11:45:09.33$& F140W	&	2.62 	&	21.7 	$\pm$	0.12 	&	14.3 	&	4.52	& 5.11 $\pm$ 0.07   &	 7.28 $\pm$ 1.88 \\
~~MACS~J1149.5$+$2223 & 0.544 	&$11:49:35.70$& $+22:23:54.68$& F160W	&	2.44 	&	34.3 	$\pm$	0.32 	&	14.3 	&	4.72	&	-			    &	 4.54 $\pm$ 1.12 \\
~~MACS~J0717.5$+$3745 & 0.548 	&$07:17:32.52$& $+37:44:34.84$& F160W	&	2.49 	&	13.2 	$\pm$	0.07 	&	14.3 	&	2.19	&	-			    &	 4.07 $\pm$ 0.73 \\
~~MACS~J0647.7$+$7015 & 0.584 	&$06:47:50.65$& $+70:14:53.99$& F160W	&	1.44 	&	56.9 	$\pm$	0.29 	&	14.3 	&  14.70	&	-			    &	 7.71 $\pm$ 2.77 \\
~~MACS~J0744.9$+$3927 & 0.686 	&$07:44:52.80$& $+39:27:26.74$& F160W	&	2.47 	&	14.7 	$\pm$	0.09 	&	14.3 	&	7.74	&	-			    &	 7.65 $\pm$ 2.45
\enddata
\tablenotetext{a}{Cluster redshift and sky coordinates.}
\tablenotetext{b}{{\em HST} band corresponding to the rest-frame
     wavelength of $1\mu$m.}
\tablenotetext{c}{S\'ersic index of the BCG obtained with GALFIT in the
     {\em HST} band corresponding to the rest-frame wavelength of $1\mu$m.}
\tablenotetext{d}{Effective radius of the BCG obtained with GALFIT in the {\em
     HST} band corresponding to the rest-frame wavelength of 1\,$\mu$m.}
\tablenotetext{e}{BCG centric radius for $\Mstar$, $\Mgas$, and $\Mtot$ estimates.}
\tablenotetext{f}{BCG total stellar mass $\Mstar(<r)$ estimated by \citet{Cooke16}.
     We assume a fractional uncertainty of $10\percent$ in our analysis.}
\tablenotetext{g}{X-ray gas mass $\Mgas(<r)$ from \citet{Donahue14}.}
\tablenotetext{h}{Lensing mass $\Mtot(<r)$ from \citet{Umetsu16}.}
\end{deluxetable*}

\section{Results}
\label{sec:results}

\begin{figure}[thb]
 \centering
 \includegraphics[width=0.48\textwidth]{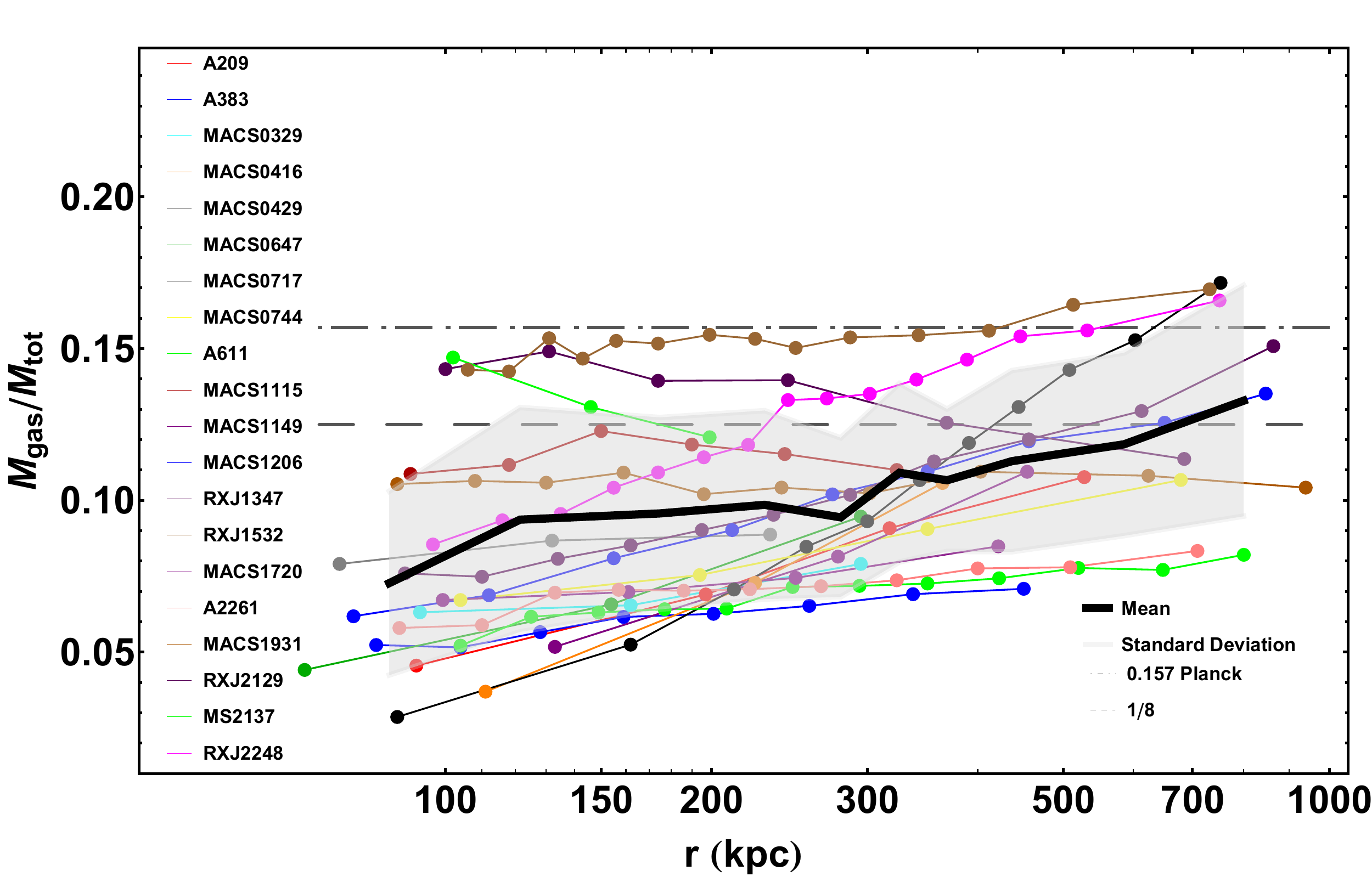}
 \includegraphics[width=0.48\textwidth]{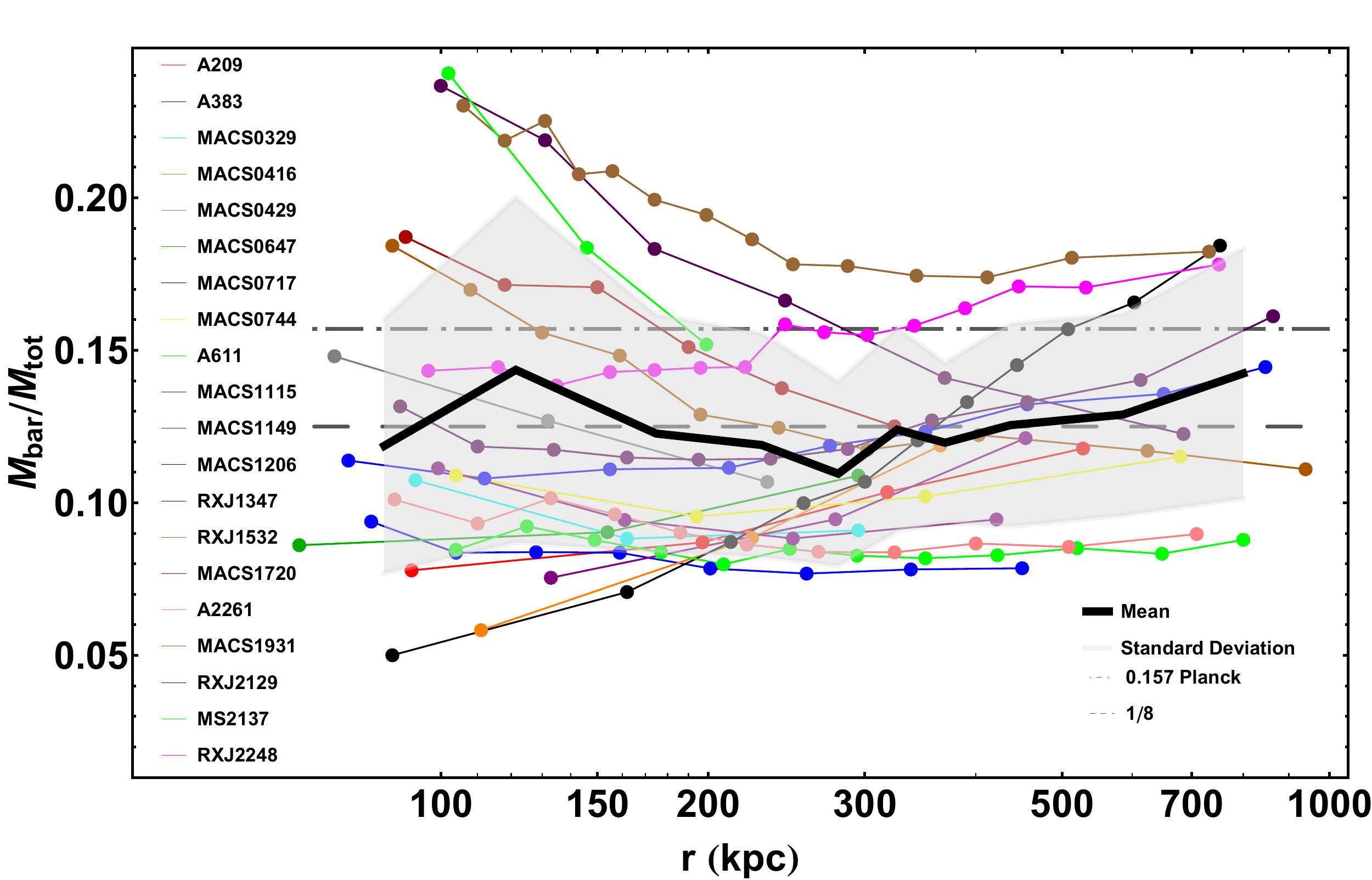}
 \caption{
Upper panel: the hot gas fraction $\fgas(r)=\Mgas(<r)/\Mtot(<r)$ as a
 function of the cluster-centric radius $r$ for 20 CLASH clusters in our sample.
Lower panel: the total baryon fraction $\fbar(r)=\Mbar(<r)/\Mtot(<r)$
 where $\Mbar$ includes the X-ray emitting hot gas mass $\Mgas$ and the
 stellar mass $\Mstar$.
 The 20 CLASH clusters are labeled with different colors.
In each panel, the thick black solid line shows the mean profile of the sample, and
 the gray shaded area represents the standard deviation from the mean.
 The horizontal dot-dashed line shows the cosmic mean baryon fraction
 $\Omega_\mathrm{b}/\Omega_\mathrm{m}$ \citep{Planck16}, and the
 horizontal dashed line corresponds to $1/8$.
 }\label{Fig:1}
\end{figure}

The key issue of the present study is to examine if the correlation
between $\gtot$ and $\gbar$ is related to any physical or environmental
parameters governing the system.
A possible approach is to study the baryon-to-total acceleration
ratio $\gbar(r)/\gtot(r)$, or the baryon fraction $\fbar(r)$,
as a function of the cluster-centric radius $r$.  An
alternative approach is to examine the correlation between $\gtot(r)$
and $\gbar(r)$, namely the RAR.

\subsection{Hot Gas Fraction}
\label{subsec:fgas}

We define the hot gas fraction as $\fgas(r) = \Mgas(<r)/\Mtot(<r)$, the
ratio of the gas mass $\Mgas(<r)$ to the total mass
$\Mtot(<r)$ as a function of the cluster-centric radius $r$.  For each
cluster, we evaluate $\Mtot(<r)$ and $\fgas(r)$ where the {\em Chandra}
gas mass measurements $\Mgas(<r)$ of \citet{Donahue14} are available.

In the upper panel of Figure~\ref{Fig:1}, we show the hot gas fractions
$\fgas(r)$ of all individual clusters in our CLASH
sample, along with the mean profile $\langle\fgas(r)\rangle$ of the
sample.
The mean $\langle\fgas(r)\rangle$ increases with increasing
cluster-centric radius $r$, approaching the cosmic baryon fraction
$\Omega_\mathrm{b}/\Omega_\mathrm{m}=(15.7\pm 0.4)\percent$
\citep{Planck16} at $r\simgt 700\,\mathrm{kpc}\sim 0.5r_{500}$.

We note that \citet{Donahue14} derived the hot gas
fractions of CLASH clusters by combining their X-ray gas mass
measurements with earlier CLASH weak-lensing results from
\citet{Umetsu14} or \citet{Merten15}. Our results improve upon those of
\citet{Donahue14} by using the full-lensing constraints of
\citet{Umetsu16} based on CLASH strong-lensing, weak-lensing shear and
magnification data.

\subsection{Baryon Fraction}
\label{subsec:fbar}

We compute for each cluster the baryon fraction
$\fbar(r)=\Mbar(<r)/\Mtot(<r)$ as a function of $r$.
The baryonic cluster mass $\Mbar(<r)$ consists of the X-ray gas mass
$\Mgas(<r)$ from \citet{Donahue14}, the stellar mass estimated as
$\Mstar(<r) = \Mgas(<r) \times \fcold(r)/[1-\fcold(r)]$
(Section~\ref{subsec:stars}), and the stellar mass of the BCG in the
innermost
cluster region (Section~\ref{subsec:BCG}).  As in
Section~\ref{subsec:fgas}, we evaluate for each cluster $\Mtot(<r)$ and
$\fbar(r)$ where the {\em Chandra} $\Mgas(<r)$ values of
\citet{Donahue14} are available.

For each cluster, we also include a single constraint on the baryon
fraction $\fbar(<r)$ in the central BCG region.
The stellar mass distribution of each BCG is modeled by Equation
(\ref{eq:Hernquist}), as described in Section~\ref{subsec:BCG}.
For 13 clusters in our sample, we have {\em Chandra} $\Mgas$
measurements \citep{Donahue14} lying in the central BCG region at
$r\ < \langle R_\mathrm{e}\rangle\sim 30$\,kpc (see Table~\ref{tab:BCG}).
For these clusters, we calculate the BCG stellar mass
$m_\mathrm{star}(<r)$
at this innermost radius of the {\em Chandra} $\Mgas$ measurements
(Table~\ref{tab:BCG}).
For the other clusters, we calculate $m_\mathrm{star}(<r)$ at
$r_\mathrm{min}\simeq 14$\,kpc and ignore the gas mass contribution to
$\Mbar(<r_\mathrm{min})$.
We note that, typically, the hot gas contribution in the innermost
region $r < R_\mathrm{e}$ is subdominant compared to the BCG stellar
mass \citep[e.g.,][]{Sartoris20}.

In the lower panel of Figure~\ref{Fig:1}, we show the baryon fraction
profiles $\fbar(r)$ outside the BCG region for all individual clusters
in our sample.
The mean $\langle\fbar(r)\rangle$ profile of the sample is nearly
constant $\sim1/8$ \citep{Donahue14} with $r$.
The cluster-to-cluster scatter around the mean $\fbar(r)$ profile is
0.041 in terms of the standard deviation, and the total baryon fractions
in some clusters reach the cosmic mean value.
We do not find any clear radial trend in the baryon fraction profiles
$\fbar(r)$ for the CLASH sample, as in the case of spiral galaxies
\citep{McGaugh04, FM12}.

At $r=800\,\mathrm{kpc}\sim 0.6r_{500}$ (the maximum radius of our
ensemble measurements), the mean baryon fraction of the
CLASH sample is $\langle\fbar\rangle=(14.2\pm 1.4)\percent$,
which corresponds to a depletion factor of
${\cal D}\equiv 1-\fbar/(\Omega_\mathrm{b}/\Omega_\mathrm{m})=(10\pm 9)\percent$
with respect to the cosmic mean value,
$\Omega_\mathrm{b}/\Omega_\mathrm{m}=(15.7\pm 0.4)\percent$.
This level of depletion is not statistically significant,
and it is in agreement with ${\cal D}= (18 \pm 2)\percent$ at
$r=r_{500}$ found from
an independent constraint on the SPT sample by \citet{Chiu18}, as well
as with the results from numerical simulations,
${\cal D}=10\percent$--$20\percent$
\citep[e.g.,][]{McCarthy11,Barnes17}.
Moreover, since the hot gas is more extended than DM, the gas fraction
$\fgas(r)$ increases with cluster-centric radius $r$,	
so that the depletion factor of the CLASH sample at $r_{500}$ is
expected to be much less significant.

\subsection{Radial Acceleration Relation in CLASH Galaxy Clusters}
\label{subsec:RAR}

\begin{figure*}[thb!]
 \centering
 \includegraphics[width=0.48\textwidth, angle=0, clip]{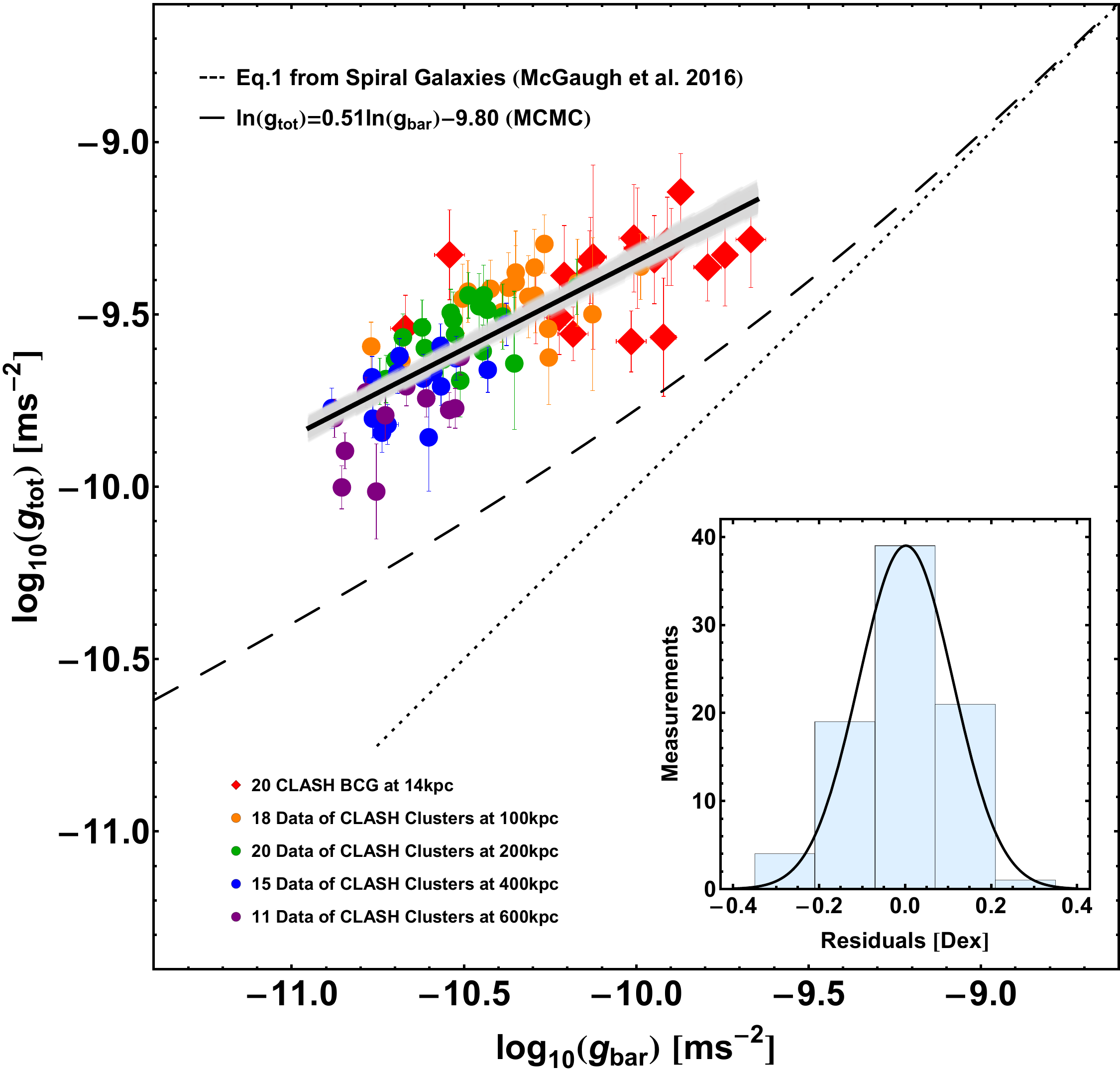}
 \includegraphics[width=0.48\textwidth, angle=0, clip]{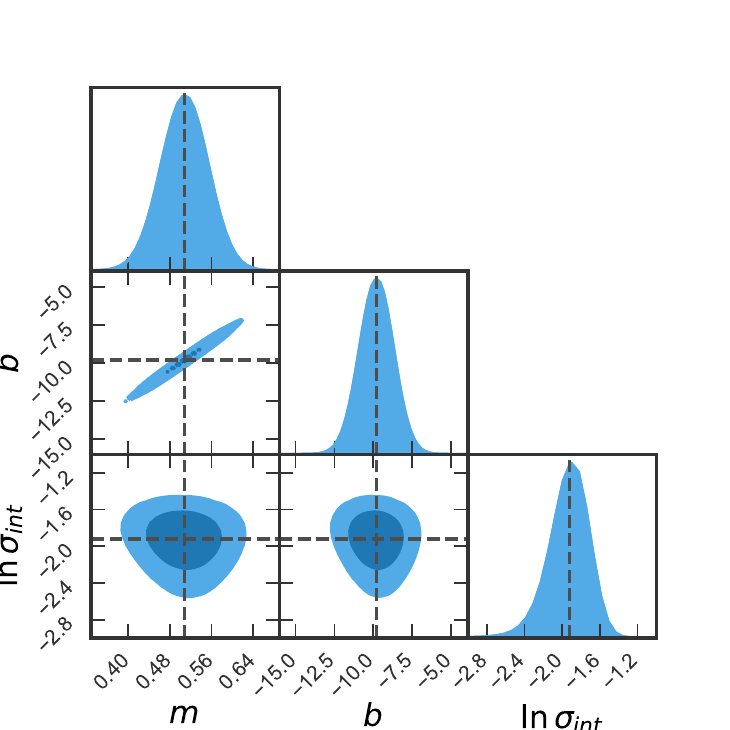}
 \caption{
Radial acceleration relation (RAR) in 20 CLASH galaxy clusters from the
 central BCG to the intra-cluster regime.
 Left panel: a comparison of the total acceleration
 $\gtot(r)=G\Mtot(<r)/r^2$ from gravitational lensing and the baryonic
 acceleration
 $\gbar(r)=G\Mbar(<r)/r^2$ which includes the hot-gas and stellar mass
 contributions to the baryonic mass $\Mbar(<r)$.
The red diamonds with error bars represent the measurements in the
 central BCGs for all 20 CLASH clusters in our sample.
The orange, green, blue, and purple circles with error bars show the
 measurements in the intra-cluster regime at $r=100, 200, 400$, and
 $600$\,kpc, respectively.
 The gray shaded area represents the $1\sigma$ range around the best-fitting
 relation shown with the black solid line.
The black dashed line shows Equation (\ref{eq:MDAR}) of \citet{McGaugh16}.
 The black dotted line shows the one-to-one relation, $\gtot=\gbar$.
 The inset plot shows the histogram distribution of best-fit residuals,
 which is characterized by a 0.11\,dex scatter.
 Right panel: constraints on the regression parameters for the RAR in
 the CLASH sample,
 showing marginalized one-dimensional (histograms)  and
 two-dimensional posterior distributions.
 }
\label{Fig:2}
\end{figure*}
	
Here we quantify and characterize the relationship between the total
acceleration $\gtot(r)$ and the baryonic acceleration $\gbar(r)$ for the
CLASH sample (Figure~\ref{Fig:1}).
To this end, for each cluster, we extract data points where possible at
$r=100$\,kpc (18 clusters), $200$\,kpc (20 clusters), $400$\, kpc
(15 clusters), and $600$\, kpc (11 clusters),
by using linear interpolation. These
data points are sufficiently well separated from each other. Hence, for
simplicity, we ignore covariances between different radial bins in our
fitting procedure.
Altogether, we have a total of 84 data points for
our sample of 20 CLASH clusters, including 20 data points in the central
BCG region.

Since the mass ratio $\Mtot(<r)/\Mbar(<r)$ is equivalent to the
acceleration ratio $\gtot(r)/\gbar(r)$,
the MDAR can be expressed as a relation between
$\gtot(r)/\gbar(r)$ and $\gbar$.
In spiral galaxies, there exists a tight empirical relationship between the
observed total acceleration $\gtot(r)$ and the baryonic acceleration
$\gbar(r)$, namely the RAR \citep{McGaugh16}.
In fact, the MDAR and RAR are mathematically equivalent
However, we point out that in the RAR, values of each of the two axes
come from independent measurements.

We present our results in the form of the RAR.
We model the CLASH data distribution in log-acceleration space
by performing a linear regression on the relation
$y=m\,x+b$ with
$y=\ln(\gtot/g_0)$ and
$x=\ln(\gbar/g_0)$ with a normalization scale of
$g_0=1\,\mathrm{m}\,\mathrm{s}^{-2}$.\footnote{We note that the
resulting RAR is presented in decimal logarithmic units ($\log_{10}$) in
Figures \ref{Fig:1} and \ref{Fig:3},
whereas our regression analysis uses the natural logarithm ($\ln$) of
acceleration.}
Here we account for the uncertainties in the determinations of
lensing mass $\Mtot(<r)$,
gas mass $\Mgas(<r)$, and
stellar mass $m_\mathrm{star}(<r)$ of the BCG
(see Section~\ref{sec:data}).
The uncertainties in $x$ and $y$
are expressed as
$\sigma_{x}=\sigma(\Mtot)/\Mtot$
and
$\sigma_{y}=\sigma(\Mbar)/\Mbar$,
where
$\sigma(\Mtot)$ and $\sigma(\Mbar)$ are
the total uncertainties for the lensing and baryonic
mass estimates $\Mtot$ and $\Mbar$, respectively.

The log-likelihood function is written as
\begin{equation}\label{eq:log-likelihood}
 -2\ln\,\mathcal{L}=\sum_{i}\,\ln{(2\pi\sigma^2_{i})}+\sum_{i}\,\frac{[y_i-(m\,x_i+b)]^2}{\sigma^2_{i}}\,,
\end{equation}
where $i$ runs over all clusters and data points, and $\sigma_i$
includes the observational uncertainties $(\sigma_{x_i}, \sigma_{y_i})$ and
lognormal intrinsic scatter $\sigma_\mathrm{int}$ \citep[e.g., see][]{Umetsu16,Okabe16},
\begin{equation}
\label{eq:sigma}
 \sigma^2_{i}=\sigma^2_{y_i}+m^2\sigma^2_{x_i}+\sigma_\mathrm{int}^2\,.
\end{equation}
The $\sigma_\mathrm{int}$ parameter accounts for the intrinsic scatter
around the mean RAR due to unaccounted astrophysics associated with
the RAR.

We perform a Markov chain Monte Carlo (MCMC) analysis to constrain the
regression parameters using the emcee python package
\citep{Foreman-Mackey13,emcee2019}
based on an affine-invariant sampler \citep{GW10}.
We use non-informative uniform priors on $b$ and $m$ of
$b\in [-100, 100]$ and $m\in [-100, 100]$. For the intrinsic scatter, we assume
a prior that is uniformed in $\ln{\sigma_\mathrm{int}}$ in the range
$\ln{\sigma_\mathrm{int}}\in [-5,1]$.
Using the MCMC technique,
we sample the posterior probability distributions of the regression
parameters $(b,m,\sigma_\mathrm{int})$ over the full parameter space
allowed by the priors.

From the regression analysis, we find a tight RAR for the CLASH sample
in the BCG--cluster regime.
Figure~\ref{Fig:2} summarizes our results.
In the left panel, we show the distribution of CLASH
clusters in $\log_{10}{\gbar}$--$\log_{10}{\gtot}$ space along with the
best-fit relation (black solid line).
The spread of the best-fit residuals is about 0.11\,dex, as shown in the
inset plot of Figure~\ref{Fig:2}.
The resulting constraints on the regression parameters are
$m=0.51^{+0.04}_{-0.05}$,
$b=-9.80^{+1.07}_{-1.08}$, and
$\sigma_\mathrm{int}=14.7^{+2.9}_{-2.8}\percent$ in terms of the
MCMC-sampled posterior mean and standard deviation.
The RAR for the CLASH sample is summarized as
\begin{equation}
 \label{eq:RAR}
 \ln(\gtot/\mathrm{m}\,\mathrm{s}^{-2})=0.51^{+0.04}_{-0.05}\,\ln(\gbar/\mathrm{m}\,\mathrm{s}^{-2})-9.80^{+1.07}_{-1.08}\,.
\end{equation}

Figure~\ref{Fig:2} also compares our results with the RAR in sprial
galaxies from \citet[][dashed line]{McGaugh16}.
The slope we obtained $m=0.51^{+0.04}_{-0.05}$ is consistent with
the low acceleration limit $1/2$ of the RAR from \citet{McGaugh16},
 $\gtot =\sqrt{g_\dag\, \gbar}$ with
 $g_\dag\simeq 1.2\times 10^{-10}$\,m\,s$^{-2}$.
On the other hand,
 the intercept
$b$ at fixed $\gbar$ is found to be significantly
 higher than that of \citet{McGaugh16}.

Here we repeat our regression analysis by fixing the slope to $m=1/2$
and rewriting the scaling relation as
$\gtot(r)=\sqrt{g_\ddag\,\gbar(r)}$ with $g_\ddag$ a constant
acceleration that corresponds to a certain characteristic acceleration
scale.
Then, the regression parameters are constrained as
 $g_\ddag = (2.02\pm0.11)\times 10^{-9}$\,m\,s$^{-2}$.
 and
 $\sigma_\mathrm{int}= 14.5^{+2.9}_{-2.8}\percent$.

\section{Discussion}
\label{sec:discussion}

\subsection{Interpretation of the CLASH RAR in the $\Lambda$CDM framework}
\label{sec:SAM}

Using a semi-analytical model with abundance matching, \citet{Navarro17}
provided a possible explanation of the RAR in spiral galaxies within the
$\Lambda$CDM framework.
Here we test if the observed RAR for the CLASH sample can be explained
by a semi-analytical description of cluster-scale halos in the standard
$\Lambda$CDM model.
Unlike the RAR in spiral galaxies based on stellar kinematics,
 the total acceleration $\gtot$ in the CLASH sample has been derived
 assuming the NFW density profile.
 Our data points and semi-analytical model are thus not entirely
 independent in terms of the profile shape of DM.

To this end, we employ the semi-analytical model of \citet{Olamaie12},
which describes the distributions of DM and hot gas
with an ideal gas equation of state
in a spherical cluster halo.
First, this model assumes that DM follows the NFW profile
\citep{Umetsu11b,Umetsu16,Niikura15,Okabe16} and
the gas pressure is described by a generalized NFW profile \citep{Nagai07}.
Following \citet{Olamaie12}, we fix the values of the gas concentration
and slope parameters of the generalized NFW profile to those found by
\citet{Arnaud10}.
Next, the system is assumed to be in hydrostatic equilibrium.
We then assume a gas mass fraction of $\fgas(r_{500})=13\percent$
\citep[e.g.,][]{planck2013fgas,Donahue14} at $r=r_{500}$ to fix the
normalization of $\rho_\mathrm{gas}(r)$.
Finally, in these calculations, we assume that
the gas density $\rho_\mathrm{gas}$ is much smaller than the DM density
$\rho_\mathrm{DM}$,
$\rho_\mathrm{tot}(r)=\rho_\mathrm{DM}(r)+\rho_\mathrm{gas}(r)\approx\rho_\mathrm{DM}(r)$.
With these assumptions, $\rho_\mathrm{gas}(r)$ can be fully specified
by two parameters that describe the NFW density profile.
We refer to \citet{Olamaie12} for full details of the model.

Following the procedure outlined above, we can describe the average
properties of our cluster sample, which includes 16 X-ray-selected and 4
high-magnification-selected CLASH clusters of \citet{Umetsu16}.
Here we adopt
$M_{200}=1.55\times10^{15} M_{\odot}$ and $c_{200}=3.28$
to describe the DM distribution $\rho_\mathrm{DM}(r)$ of our
sample with a median redshift of $z=0.377$.
Given the NFW parameters, we compute the gas density profile
 $\rho_\mathrm{gas}(r)$ for the CLASH sample.
We use $\fcold(r)$ of \citet{Chiu18} (Section~\ref{subsec:stars}) to
 account for the stellar mass contribution to the baryonic mass.
With these average profiles $\rho_\mathrm{DM}(r)$,
 $\rho_\mathrm{gas}(r)$, and $\fcold(r)$,
 we can predict the total and baryonic
 gravitational acceleration profiles, $\gtot(r)$ and $\gbar(r)$, for the
 CLASH sample.  Here we
 compute the acceleration profiles at $r=100, 200, 400$, and $600$\,kpc,
 as done in our CLASH analysis.

\begin{figure}[thb!]
 \centering
 \includegraphics[width=0.47\textwidth]{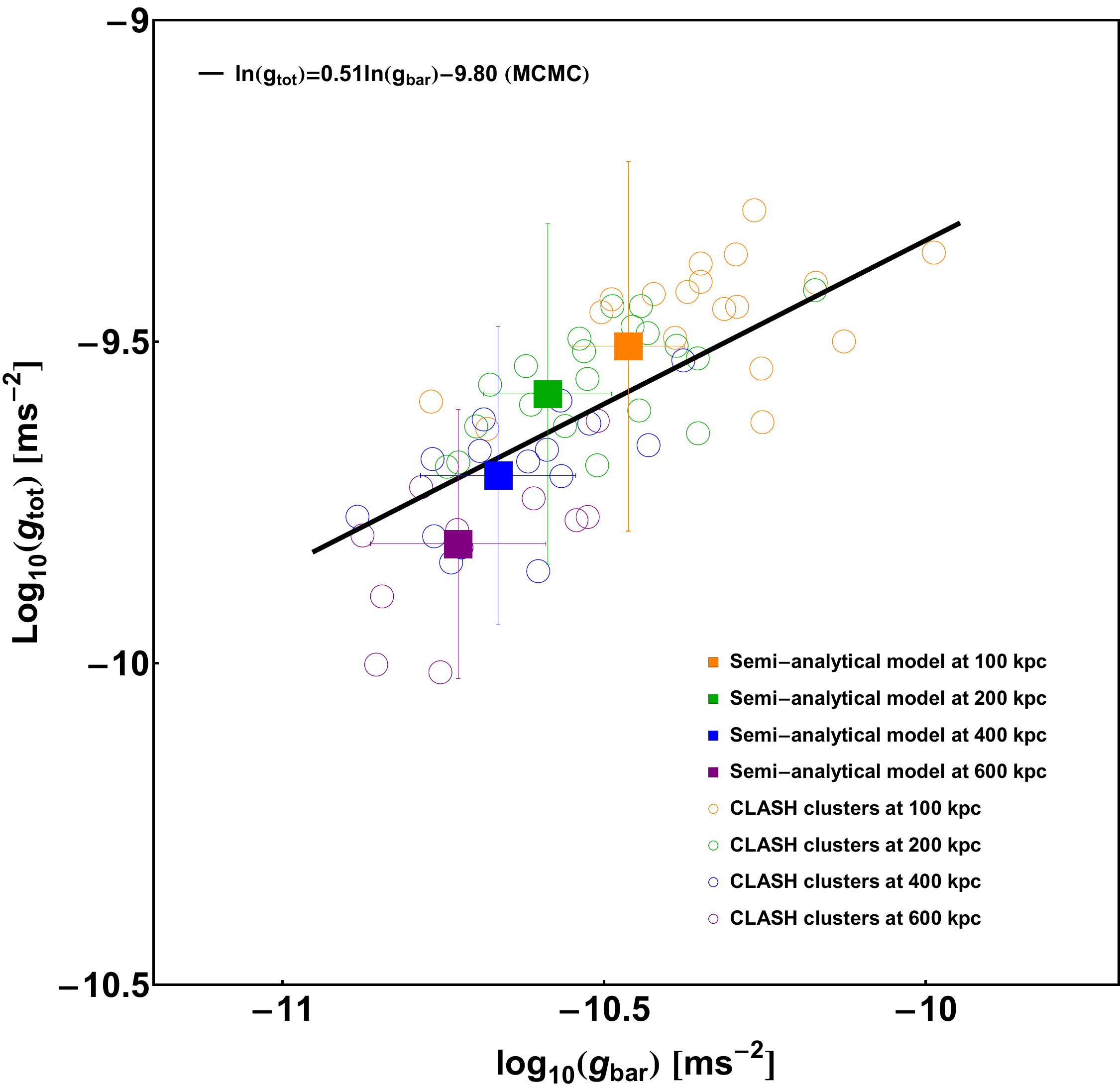}
 \caption{
Predictions for the CLASH radial acceleration relation (RAR) from
  semi-analytic modeling in the $\Lambda$CDM framework.
 The orange, green, blue, and purple squares
 show the model predictions for the CLASH sample at $r = 100, 200, 400$, and $600$\,kpc,
 respectively, and the error bars represent the $1\sigma$ intrinsic scatter.
 Open circles show the measurements for the CLASH sample (Figure~\ref{Fig:2}).
}\label{Fig:3}
\end{figure}

We also model the intrinsic scatter around the average profiles
 $\gtot(r)$ and $\gbar(r)$ due to cluster-to-cluster variations in the
 DM and baryonic distributions.
For the DM distribution, we assign intrinsic scatter in
 $c_{200}$ with a lognormal intrinsic dispersion of $30\%$
 \citep[e.g.,][]{Bhattacharya13}.
For the baryonic distribution, we assign intrinsic scatter in $\fcold$
 with a Gaussian dispersion of $12\%$ \citep{Chiu18}.
We employ Monte-Carlo simulations to evaluate the intrinsic dispersions
 around the average acceleration profiles $\gtot(r)$ and $\gbar(r)$ at
 each cluster-centric radius.
The result is shown in Figure~\ref{Fig:3}.

The 20 galaxy clusters in our sample are all high-mass systems selected
for the CLASH survey \citep{Postman12}.
It is thus reasonable to adopt a set of average properties in the
semi-analytical model to study the new RAR.
As shown in Figure~\ref{Fig:3}, all the model points (color-coded
squares) match the distribution of data points (color open circles) and
the mean relation (black solid line) well.

The inferred level of $1\sigma$ intrinsic scatter estimated by
Monte-Carlo simulations appears to be larger than the data
distribution.
This is similar to the findings in spiral galaxies \citep[see,
e.g.,][]{DL16, Desmond17, Lelli17}.
However, it should be noted that the CLASH sample is
dominated by relaxed systems \citep{Meneghetti14}
because of the CLASH selection based on X-ray morphology
regularity \citep{Postman12}. As a result, the CLASH sample is predicted
to have a much smaller level of intrinsic scatter in the
$c_{200}$--$M_{200}$ relation \citep[$16\percent$;][]{Meneghetti14,Umetsu16}.

The CLASH RAR (Equation (\ref{eq:RAR})) expresses $\gtot(r)$ as a
function of $\gbar(r)$.
Thus, one can obtain the baryon fraction $\fbar(r)$ if $\gbar(r)$ is
known, because $\fbar(r)=\Mbar(<r)/\Mtot(<r)=\gbar(r)/\gtot(r)$.
If we approximate the CLASH RAR as $\gtot\approx\sqrt{g_\ddag\,\gbar}$,
then the baryon fraction has the simple form,
\begin{equation}
 \label{eq:fbar_gbar}
 \fbar(r)\approx\sqrt{\gbar(r)/g_\ddag}.
\end{equation}

Since the 20 CLASH clusters in our sample are of similar size and
acceleration profile, we can infer an average relation between the
baryonic acceleration $\gbar$ and the cluster-centric radius $r$.
Making use of the X-ray gas mass measurements from \citet{Donahue14} and the
stellar mass correction from \citet{Chiu18},
we obtain such an average relation, i.e., the average baryonic
acceleration as a function of $r$, $\langle\gbar(r)\rangle$.
Then, the best-fit CLASH RAR (Equation (\ref{eq:RAR})) together with
$\langle\gbar(r)\rangle$
gives an empirical relation for the baryon fraction as a function of
$r$.

\begin{figure}[thb!]
 \centering
 \includegraphics[width=0.47\textwidth]{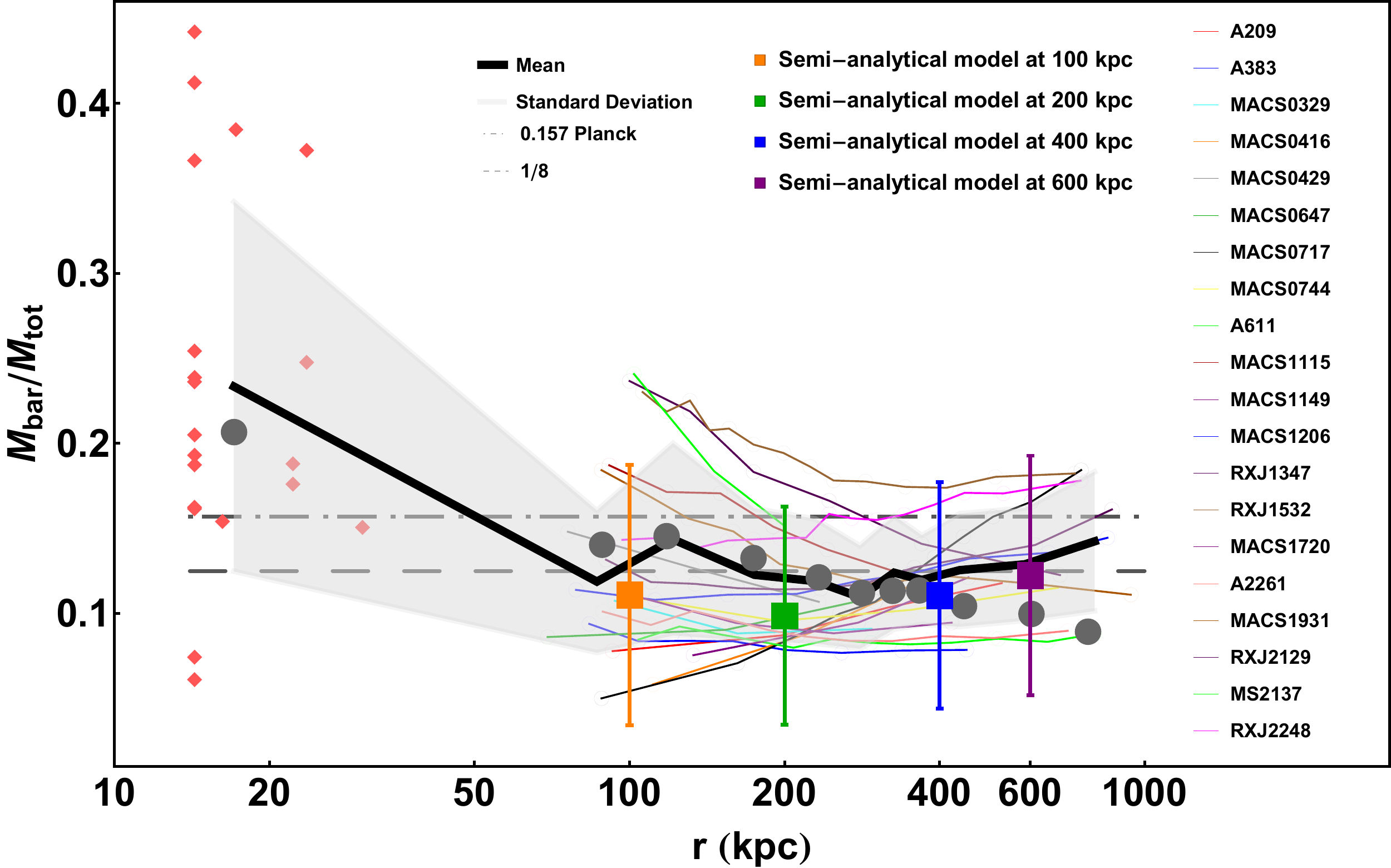}
 \caption{
Comparison of the observed and predicted total baryon fraction profiles
 $\fbar(r)=\Mbar(<r)/\Mtot(<r)$ for the CLASH sample.
The red diamonds and color-coded lines represent the CLASH
 observational constraints in the BCG and cluster regions, respectively.
 The gray circles show empirical predictions
 based on the best-fit CLASH RAR (see Equation (\ref{eq:RAR})) combined
 with the mean value of baryonic acceleration
 $\langle\gbar(r)\rangle$ at each cluster radius.
The thick black line shows the mean $\langle\fbar(r)\rangle$ profile
 averaged over the CLASH sample.
The orange, green, blue, and purple squares represent
 predictions from a semi-analytical model
 at $100, 200, 400$, and $600$\,kpc respectively,
 and the error bars show the $1\sigma$
 intrinsic scatter.
 }\label{Fig:4}
\end{figure}

In Figure~\ref{Fig:4}, we compare the $\fbar(r)$ profile inferred from
the best-fit CLASH RAR (Equation (\ref{eq:RAR})) with the observed
distribution of CLASH baryon fractions shown in Figure~\ref{Fig:1}.
The gray circles are the inferred baryon fractions at different
cluster-centric radii, $r$.
The thick black line shows the observed mean $\langle\fbar(r)\rangle$
profile averaged over the CLASH sample (see Figure~\ref{Fig:1}).
The $\fbar(r)$ profile inferred from the best-fit RAR and
the mean $\langle\fbar(r)\rangle$ profile for the CLASH sample
are consistent within $r=400$\,kpc, but deviate from each other beyond
$400$\,kpc.
This is because the best-fit RAR model predicts a radially
decreasing $\fbar(r)$ profile, whereas the observed CLASH
$\langle\fbar(r)\rangle$ profile is nearly constant ($\sim 1/8$; see
Figure~\ref{Fig:1}) in the intracluster regime.

We also compare these baryon fraction profiles with
the corresponding values predicted by our semi-analytical model,
which are denoted by the orange, green, blue and purple squares
for $r=100, 200, 400$, and $600$\,kpc, respectively.
The values of all model points agree well with the best-fit RAR and the
observed mean values for the CLASH sample.
Our semi-analytical model predicts a nearly constant $\fbar(r)$ profile,
in good agreement with the mean relation of the CLASH sample (black
solid line).

\begin{figure}[thb!]
 \centering
 \includegraphics[width=0.47\textwidth]{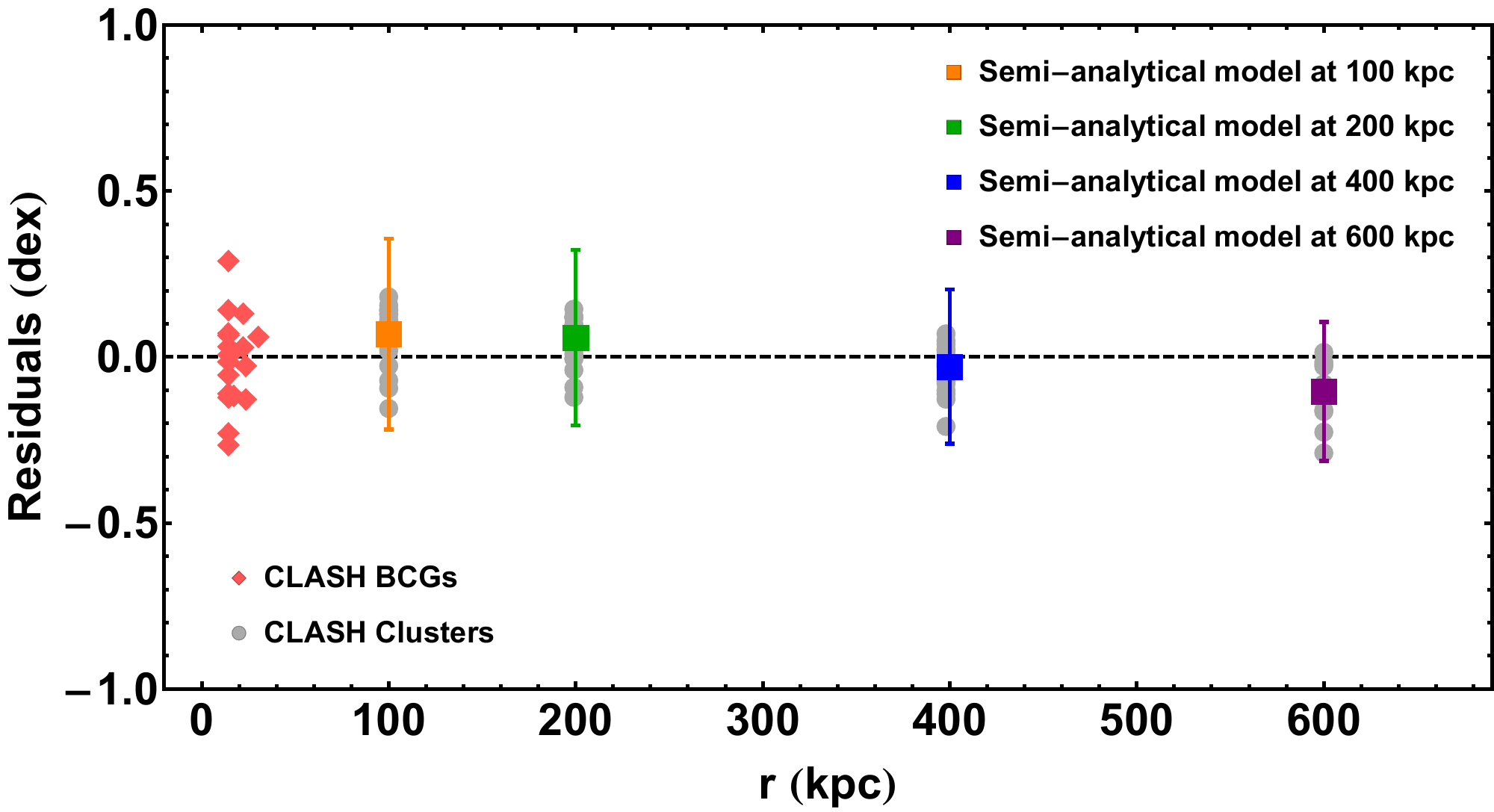}
 \caption{
Residuals of the best-fit RAR for the CLASH sample as a function of the
 cluster-centric radius $ r$.
The vertical axis represents the difference between the data and the
 best-fit RAR in units of dex.
The horizontal axis is the cluster-centric radius $r$ in units of kpc.
The red diamonds and the gray filled circles denote the CLASH
 measurements in the central BCG region and the intracluster region
 ($100, 200, 400$, and $600$\,kpc), respectively (see Figure~\ref{Fig:2}).
The orange, green, blue, and purple squares show the
 deviations of our semi-analytical predictions
 relative to the best-fit CLASH RAR, and the error bars show the $1\sigma$
 intrinsic scatter.
}\label{Fig:5}
\end{figure}

To assess how the data deviate from the best-fit CLASH RAR,
we show in Figure~\ref{Fig:5} the distribution of best-fit residuals as
a function of cluster radius $r$.
We note again that both the semi-analytical model and the CLASH
RAR assume the NFW density profile for the total matter distribution.
The residuals are defined as the difference of $\log_{10}(\gtot)$ between
the observational data and the best-fit RAR (Equation (\ref{eq:RAR})).
As shown in the figure, the residuals at small $r$ ($\approx14$, 100,
and 200\,kpc) distribute symmetrically across the best-fit RAR (the zero
residual line).
However, at larger cluster radii $r\simgt 400$\,kpc, the residuals deviate
systemically toward negative values.
The semi-analytical model (color-coded squares) exhibits a similar trend
as seen for the observational data.

\subsection{Implications for Residual Missing Mass in MOND}

As discussed earlier, Equation (\ref{eq:MDAR}) in the MOND framework is
consistent with the observed RAR on galaxy scales with $\gbar\equiv
g_\mathrm{M}$, without introducing DM.
Here we test this hypothesis with our CLASH RAR results and infer the level
of residual missing mass on the BCG-cluster scale within the MOND
framework.

Since the total acceleration $\gtot$ in the CLASH RAR implies
 $\gtot\approx\sqrt{\gbar\,g_\ddag}$,
 we can relate $\gbar$ and $g_\mathrm{M}$ by
\begin{equation}\label{eq:missing baryon}
\sqrt{\gbar\,g_\ddag}\approx \frac{g_{\mathrm{M}}}{1-e^{-\sqrt{g_{\mathrm{M}/g_\dag}}}}\,.
\end{equation}
The mass ratio between $M_\mathrm{M}$ and $\Mbar$ is
$M_{\mathrm{M}}/\Mbar=g_{\mathrm{M}}/\gbar$.
Our formulation implies that the level of residual missing mass in MOND
depends on the baryonic acceleration $\gbar$.
In our CLASH data,
the largest baryonic acceleration in the BCG regime
is $2.1\times10^{-10}$\,m\,s$^{-2}$ and
 the smallest one in the intracluster regime is
 $1.3\times10^{-11}$\,m\,s$^{-2}$.
The corresponding mass ratio $M_{\mathrm{M}}/\Mbar$ thus ranges from
$2.7$ to $7.3$, increasing with decreasing baryonic acceleration $\gbar$
(or increasing cluster-centric radius).

 Hence,
 the CLASH RAR confirms the existence of residual missing mass in MOND
 on the BCG-cluster scale,
 as found in the literature \citep{Sanders99, Sanders03, PS05, FM12}.
 Furthermore, it reveals a more substantial level of
 discrepancy or residual missing mass $M_\mathrm{M}/\Mbar$ in the low
 $\gbar$ regime, which corresponding to the cluster outskirts.
 The typical level of residual missing mass found in the literature is
 $\langle M_\mathrm{M}/\Mbar\rangle\sim 2$.
However, it should be note that a fair comparison of the CLASH RAR with
MOND will require a relativistic extension of MOND to properly interpret
the gravitational lensing data for the CLASH sample.
Some possibilities in this approach for cluster lensing have been discussed
 in the literature \citep{Bruneton09, ZF12, FM12}.

\subsection{Implications for Kinematic Scaling Relations}

We recall that the RAR in galaxies has  a characteristic acceleration
scale of $g_\dag\simeq 1.2\times 10^{-10}$\,m\,s$^{-2}$,
above which the acceleration of the system asymptotically tends to
Newtonian dynamics without DM (i.e., $\gtot=\gbar$),
while below which the acceleration asymptotically tends to
$\gtot=\sqrt{g_\dag\, \gbar}$.
Our CLASH results indicate that $\gtot\approx\sqrt{g_\ddag\, \gbar}$
(see Equation (\ref{eq:fbar_gbar})) on BCG--cluster scales, with a
characteristic acceleration scale of $g_\ddag \gg g_\dag$.
However, in our CLASH sample, it is not clear whether or not the
acceleration of the system will approach to $\gtot=\gbar$
in the high acceleration limit, $\gbar\gg g_\ddag$.

If the CLASH RAR holds in general for other galaxy clusters,
then there exists a kinematic relation or law in galaxy clusters,
in analogy to Kepler's law of planetary motion that comes from Newtonian dynamics.
Let us take the example of the baryonic Tully--Fisher relation
\citep[BTFR; e.g.,][]{McGaugh11, McGaugh12, FM12},
 which comes from the low acceleration end of the RAR of spiral galaxies.
In rotationally support systems such as spiral galaxies, the centripetal
acceleration is provided by the total gravitational acceleration,
i.e., $v^2/r=\gtot$ where $v$ is the circular speed.
In the small acceleration regime of $\gbar\ll g_\dag$,
the RAR in galaxies gives $\gtot\approx\sqrt{g_\dag\, \gbar}$.
By expressing the baryonic acceleration as $\gbar=G\Mbar/r^2$
where $\Mbar$ is the total baryonic mass inside $r$,
we obtain the BTFR, $v^4=G g_{\dag}\Mbar$.
For systems supported by random motions, such as elliptical galaxies and galaxy
clusters, $\gtot\propto\sigma^2/r$ with $\sigma_v$ the velocity dispersion.
If a system that follows the RAR is in the low acceleration regime,
$\sigma_v^4\propto g_{\ddag}\Mbar$ is anticipated.
We refer to this kinematic law as the baryonic Faber--Jackson relation
(BFJR).

In the literature, the BFJR has not been confirmed in galaxy
clusters.  However, scaling relations between total cluster mass and
X-ray mass proxies \citep[e.g., X-ray gas temperature;][]{Sanders94,
Ettori04, Angus08, FM12}, and that between X-ray luminosity and galaxy
velocity dispersion \citep{XW00, Sanders10, Zhang11, Nastasi14}, have
been firmly established based on multiwavelength observations and
numerical simulations.
If the total baryonic mass in galaxy clusters is
tightly coupled with thermodynamic properties of the hot gas in
hydrostatic equilibrium, we may expect a correlation between
the total baryonic mass and galaxy velocity dispersion.

\section{Summary}
\label{sec:summary}

The radial acceleration relation (RAR) in galaxies represents a tight
 empirical scaling law between the total acceleration
 $\gtot(r)=G\Mtot(<r)/r^2$
 observed in galaxies and that expected from their baryonic mass
 $\gbar(r)=G\Mbar(<r)/r^2$,
 with a characteristic acceleration scale of
 $g_\dag\simeq 1.2\times 10^{-10}$\,m\,s$^{-2}$ \citep{McGaugh16}.
The RAR observed on galaxy scales raised four fundamental issues
 to be explained \citep[see Section~\ref{sec:intro};][]{Desmond17, Lelli17}.

In this paper, we have examined if such a correlation exists in galaxy
 clusters using weak-lensing, strong-lensing, and X-ray data sets
 \citep{Donahue14,Umetsu16} available for 20
 high-mass clusters targeted by the CLASH survey \citep{Postman12}.
By combining our CLASH data sets with central baryonic mass in the BCG
region and accounting for the stellar baryonic component in the
intracluster region, we have discovered, for the first time, a tight RAR
on BCG--cluster scales.
The resulting RAR for the CLASH sample is well described by a power-law
 relation,
 $\gtot\propto g_\mathrm{bar}^{0.51^{+0.04}_{-0.05}}$, with
 lognormal intrinsic scatter of $14.7^{+2.9}_{-2.8}\percent$.
The slope of the best-fit relation is consistent with the low
acceleration limit of the RAR in galaxies,
$\gtot =\sqrt{g_\dag\, g_\mathrm{bar}}$,
 whereas the intercept implies a much higher acceleration scale
 of $g_\ddag = (2.02\pm0.11)\times 10^{-9}$\,m\,s$^{-2}$.
Our results indicate that there is no universal RAR that holds on all
scales from galaxies to clusters.

Regarding the issues raised by the RAR in galaxies,
the CLASH RAR has:
(1) an acceleration scale $g_\ddag$ that is much higher than that in
galaxies, $g_\ddag \gg g_\dag$;
(2) the slope in the best-fit RAR is $0.51^{+0.04}_{-0.05}$,
which matches the low acceleration limit of the RAR in galaxies (see
Equation (\ref{eq:MDARlow}));
(3) the level of intrinsic scatter, $14.7^{+2.9}_{-2.8}\percent$,
is as tight as that in the RAR for galaxies.
The best-fit residuals of the CLASH RAR exhibit a systematic radial
trend at $r>400$\,kpc (Figure~\ref{Fig:5}).
The best-fit RAR model predicts a radially decreasing $\fbar(r)$
profile, whereas the CLASH data distribution is nearly constant
(Figure~\ref{Fig:1}) in the intracluster regime.
To fully investigate the discrepancy at
$\gbar\simlt \times10^{-11}$\,m\,s$^{-2}$, or at $r\simgt 400$\,kpc,
we need additional data covering a broader range of acceleration on
BCG--cluster scales.

We find that the observed RAR in CLASH clusters is consistent with
predictions from semi-analytical modeling of cluster halos in the
standard $\Lambda$CDM framework.
Our results also predict the presence of a baryonic Faber--Jackson
relation ($\sigma_v^4\propto M_\mathrm{bar}$) on cluster scales.


\acknowledgments

We are very grateful to Stacy McGaugh for stimulating discussions on this work.
We thank Po-Chieh Yu for his help with the GALFIT analysis.
We thank the anonymous referee for their valuable comments to improve
the clarity of this paper.
Y.T. and C.M.K. are supported by the Taiwan Ministry of Science and
Technology grant MOST 105-2112-M-008-011-MY3 and MOST 108-2112-M-008-006.
K.U. acknowledges support from the Ministry of Science and Technology of
Taiwan (grant MOST 106-2628-M-001-003-MY3) and from
the Academia Sinica Investigator Award (AS-IA-107-M01).
We acknowledge the use of the pyGTC package \citep{Bocquet16} for
creating the right panel of Figure \ref{Fig:2}.\\

\noindent
\textit{Note added in proof.}
We noted that while this paper was under review for publication, a paper
 by \cite{CD16} appeared on the arXiv preprint service.
They analyzed X-ray data for a sample of X-ray-selected non-cool-core
 clusters and derived an RAR for their sample assuming hydrostatic
 equilibrium, without accounting for the stellar baryonic contribution.
Although their results are not in quantitative agreement with ours, they
 also conclude that the RAR is unlikely to be universal and scale
 independent.

\bibliographystyle{yahapj}
\bibliography{reference}

\end{document}